\documentclass[prb,letterpaper,twocolumn,aps,floatfix]{revtex4-1}
\usepackage{graphicx}
\usepackage{xcolor}
\usepackage{amsmath}
\usepackage{amsfonts}
\usepackage[small,bf]{subfigure}
\newcommand{\beq}{\begin{equation}}
\newcommand{\eeq}{\end{equation}}
\newcommand{\bk}{{{\bf{k}}}}

\newcommand{\br}{{{\bf{r}}}}
\newcommand{\bR}{{{\bf{R}}}}

\newcommand{\bb}{{\bf{b}}}

\newcommand{\beqa}{\begin{eqnarray}}
\newcommand{\eeqa}{\end{eqnarray}}
\newcommand{\pdg}{{\vphantom \dag}}
\newcommand{\dg}{{\dag}}

\newcommand{\ra}{\rightarrow}

\newcommand{\cL}{{\cal L}}
\begin{document}
\title{Topological order in interacting semimetals}
\author{Jinmin Yi}
\thanks{These authors contributed equally to this work.}
\affiliation{Department of Physics and Astronomy, University of Waterloo, Waterloo, Ontario 
N2L 3G1, Canada} 
\affiliation{Perimeter Institute for Theoretical Physics, Waterloo, Ontario N2L 2Y5, Canada}
\author{Xuzhe Ying}
\thanks{These authors contributed equally to this work.}
\affiliation{Department of Physics and Astronomy, University of Waterloo, Waterloo, Ontario 
N2L 3G1, Canada} 
\affiliation{Perimeter Institute for Theoretical Physics, Waterloo, Ontario N2L 2Y5, Canada}
\author{Lei Gioia}
\affiliation{Department of Physics and Astronomy, University of Waterloo, Waterloo, Ontario 
N2L 3G1, Canada} 
\affiliation{Perimeter Institute for Theoretical Physics, Waterloo, Ontario N2L 2Y5, Canada}
\author{A.A. Burkov}
\affiliation{Department of Physics and Astronomy, University of Waterloo, Waterloo, Ontario 
N2L 3G1, Canada} 
\affiliation{Perimeter Institute for Theoretical Physics, Waterloo, Ontario N2L 2Y5, Canada}
\date{\today}
\begin{abstract}
It has recently been demonstrated that it is possible to open a gap in a magnetic Weyl semimetal, while preserving the chiral anomaly along with 
the charge conservation and translational symmetries, which all protect the gapless nodes in a weakly interacting semimetal.
The resulting state was shown to be a nontrivial generalization of a nonabelian fractional quantum Hall liquid to three dimensions. 
Here we point out that a second fractional quantum Hall state exists in this case. This state has exactly the same electrical and thermal Hall responses 
as the first, but a distinct (fracton) topological order. 
Moreover, the existence of this second fractional quantum Hall state necessarily implies a gapless phase, which has identical topological response to a noninteracting Weyl semimetal, but is distinct from it. This may be viewed as a generalization (in a weaker form) of the known duality between a noninteracting two-dimensional Dirac fermion and QED$_3$ to $3+1$ dimensions. In addition we discuss a $(3+1)$-dimensional topologically ordered state, obtained by gapping a nodal line semimetal without breaking symmetries. 
\end{abstract}
\maketitle
\section{Introduction}
\label{sec:1}
Topological order, a concept that originated in the study of the fractional quantum Hall effect (FQHE) in two dimensional (2D) electron gas systems,~\cite{Wen90} continues to be a subject of intense interest. 
From the fundamental physics prospective, topologically ordered states provide perfect examples of emergent macroscopic quantum phenomena, 
with fractionally-quantized electromagnetic and thermal responses, that are impossible to explain based on textbook models of weakly-interacting 
electrons. Instead, such fractionally-quantized observable responses necessarily imply excitations with fractional charges, fractional and nonabelian 
statistics, which can not be constructed out of any finite number of elementary constituents.~\cite{Wen_QFT}
In addition, such exotic excitations may have future potential practical uses in quantum computing and quantum simulation, as their
nonlocal topological nature makes them highly resistant to decoherence and noise.~\cite{RMP_QC}

Topologically-ordered states in 2D are by now well-understood. 
Various theoretical models,~\cite{Laughlin83,Zhang92,Levin05,Kitaev06} as well as complete formal classifications of 2D topological orders exist.~\cite{Wen17}
Although significant progress has been made in recent years,~\cite{Levin-Stern,Maciejko10,Swingle11,Walker-Wang,Burnell13,Qi14,Wang-Levin,YeGu15,YeGu16,Furusaki19,Lan19} less is known about topologically-ordered states in three dimensions (3D). 
3D topologically ordered states are significantly different from the 2D ones. On the one hand, fractional statistics is impossible in 3D and quasiparticle 
excitations may only be bosons or fermions. This could make one doubt that, for example, fractional quantum Hall (FQH) states may even in principle 
be generalized to 3D, as the existence of anyons, i.e. quasiparticles with fractional statistics, is an essential feature of the 2D FQHE.  On the other hand, in addition to point quasiparticles, one-dimensional loop excitations exist in 3D, which both adds 
complexity and opens up new interesting possibilities. 

We recently demonstrated that a promising way to achieve 3D topologically ordered states is through gapping topological semimetals without breaking 
the protecting symmetries~\cite{Wang20,Thakurathi20,Sehayek20} (see Refs.~\onlinecite{Meng16,Morimoto16,Sagi18,Kane18,Meng19,Teo19,Kane22} for related work).
Topological semimetals~\cite{Volovik03,Volovik07,Murakami07,Wan11,Burkov11-1,Weyl_RMP} are intermediate phases between insulators of different electronic structure topology.
They may be characterized by unquantized anomalies,~\cite{Gioia21,Wang21} i.e. topological terms with noninteger and 
continuously-tunable coefficients, similar to the electron filling parameter, characterizing ordinary Fermi liquids. 
Much like fractional filling in a Fermi liquid mandates the existence of a Fermi surface of gapless particle-hole excitations,~\cite{Else21}
these unquantized anomalies necessarily imply gapless modes and corresponding long-range entanglement. 
The only way gaplessness may be circumvented in the absence of broken symmetries is through the formation of a topologically-ordered state, 
which preserves the anomaly and the long-range entanglement of the gapless semimetal. 

Specifically, in Ref.~\onlinecite{Wang20} we presented an explicit construction of a 3D topologically-ordered state in a gapped magnetic Weyl 
semimetal, which exhibits a nontrivial generalization of the FQHE to 3D. 
This state is obtained starting from a magnetic Weyl semimetal with a single pair of nodes, separated by half a reciprocal lattice vector. 
These nodes may be gapped by breaking the $U(1)$ charge conservation symmetry while forming a superconducting state with intra-nodal pairing. 
In general, such states break translational symmetry since the Weyl nodes exist at nontrivial momenta in the first Brillouin zone (BZ). 
However, when the nodes are separated by exactly half a reciprocal lattice vector, such a pairing leads to density modulation at the reciprocal lattice vector, 
which does not break the crystal translational symmetry. 
Restoring the charge conservation symmetry by proliferating flux $2 h c/ e = 4 \pi$ (we will be using $\hbar = c = e = 1$ units throughout this paper) 
vortices in the superconducting order parameter leads to a featureless fractionalized insulator with $\mathbb{Z}_4$ topological order, that has 
the same electrical and thermal Hall conductivities as the original noninteracting Weyl semimetal, i.e. exhibits FQHE in 3D. 
Unlike in 2D FQH liquids, quasiparticle excitations in this state are bosons and fermions. What plays the role of the anyons in the 2D FQHE are 
intersections of the vortex-loop excitations with atomic planes. These behave as fractionally-charged particles with semionic statistics, which may 
be sharply defined by considering three-loop braiding processes,~\cite{Wang-Levin} involving a line defect of translational symmetry, i.e. an edge dislocation. 

In this paper we show that, in addition to the 3D FQH state of Ref.~\onlinecite{Wang20}, another state exists, which has identical topological 
response, but distinct topological order, which turns out to be of a fracton type. 
The existence of these two distinct states turns out to be closely related to a very similar property of gapped symmetric 2D Dirac surface states of 3D 
time-reversal (TR) invariant topological insulators (TI).~\cite{Bonderson13,Wang13,Fidkowski14,Metlitski15,Mross15}
In this case, two distinct topologically-ordered states exist. One, called Pfaffian-antisemion,~\cite{Wang13,Metlitski15} is closely 
related to the 3D FQH states of Ref.~\onlinecite{Wang20} (more precisely, the relation is with the TR-broken version of this state).
The second one, T-Pfaffian,~\cite{Bonderson13,Fidkowski14} is related to the other 3D state we will construct in the present paper (again, 
more precisely, the relation is with the TR-breaking version of this state, which is usually called PH, which stands for particle-hole-symmetric, 
Pfaffian~\cite{Son15,Son18}). 

Another interesting consequence emerges from these analogies to the 2D TR-invariant TI surface state topological orders. It is well-known that the PH-Pfaffian  is closely related to the recently discovered duality relation between a massless noninteracting 
2D Dirac fermion and QED$_3$.~\cite{Wang-Senthil,Metlitski-Vishwanath,Son15,Alicea16,Seiberg16,Karch16,Raghu18} 
Namely, the PH-Pfaffian state is obtained when the dual Dirac fermion of QED$_3$ is gapped by pairing, which does not break the charge
conservation symmetry since the dual fermion is neutral. 
The existence of the analog of the PH-Pfaffian state in our 3D system then also implies the existence of a gapless state, which is related 
to the noninteracting Weyl semimetal via a duality relation, somewhat similar to the 2D Dirac duality. 
We demonstrate that this is indeed the case. However, we find that the duality only applies to topological response in this case and not to the dynamics and is weaker than the 2D duality in this sense. 

The path to topologically ordered insulators through gapping topological semimetals is quite general and is not limited to the magnetic Weyl semimetal case.
To emphasize this point, here we also discuss a topologically-ordered state, which is obtained by gapping a nodal line semimetal without breaking symmetries. This state has a topological order, distinct from a gapped Weyl semimetal, and is characterized by a fractional electric polarization, impossible in an ordinary weakly-interacting insulator. 

The rest of the paper is organized as follows. 
In section~\ref{sec:2}, after a preliminary discussion of topological field theory description of the electromagnetic response of Weyl 
semimetals, we recap the construction of the 3D analog of the Pfaffian-antisemion state of Refs.~\onlinecite{Wang20,Thakurathi20}.
In section~\ref{sec:3}, we demonstrate the existence of a duality relation (which applies to topological response only) between a noninteracting Weyl semimetal and a QED$_4$, which describes a time-reversed Weyl semimetal, coupled to a dynamical gauge field. 
In section~\ref{sec:4} we discuss a topologically-ordered state, obtained by gapping a nodal line semimetal without breaking symmetries. 
This state is characterized by a fractional electric polarization, impossible in an ordinary insulator. 
We conclude in section~\ref{sec:5} with a brief discussion of our results. 

\section{Gapped symmetry-preserving states in Weyl semimetals}
\label{sec:2}
\subsection{Preliminaries}
\label{sec:2.1}
To keep the paper self-contained, we will start by recapping the construction of the 3D FQH state of Ref.~\onlinecite{Wang20,Thakurathi20}, 
which, as will be explained below, may be viewed as a TR-breaking 3D analog of the Pfaffian-antisemion state on a strongly-interacting 3D 
TI surface. We will also put the theory of Ref.~\onlinecite{Thakurathi20} on a more rigorous footing by introducing the language of translation gauge fields,~\cite{Volovik19,Nissinen21,Song21,Barkeshli21,Gioia21}
which allows one to use proper coordinate-free notation for the corresponding topological field theories. 

We start from the simplest cubic lattice model of a magnetic Weyl semimetal with a pair of nodes~\cite{Burkov11-1}
\beq
\label{eq:1}
H = \sum_{\bk} \psi^\dg_\bk \left[\sigma_x \sin(k_x d) + \sigma_y \sin(k_y d) + \sigma_z m(\bk)\right] \psi^\pdg_\bk. 
\eeq
Here $\sigma_i$ are Pauli matrices, describing the pair of touching bands and
\beq
\label{eq:2}
m(\bk) = \cos(k_z d) - \cos(Q d) - \tilde m [2 - \cos(k_x d) - \cos(k_y d)], 
\eeq
where $d$ is the lattice constant, $\tilde m > 1$ and $m(\bk)$ vanishes at two points on the $z$-axis with $k_z = \pm Q$, which correspond to the locations of the Weyl nodes.

Such a Weyl semimetal is characterized by the anomalous Hall conductivity
\beq
\label{eq:3}
\sigma_{xy} = \frac{e^2}{h} \frac{2 Q}{2 \pi} = \frac{1}{2 \pi}\frac{2 Q}{2 \pi}. 
\eeq
This may be expressed as a topological term in the effective action for probe electromagnetic gauge fields when the fermions are 
integrated out~\cite{Zyuzin12-1}
\beq
\label{eq:4}
\cL = i \frac{2 Q}{8 \pi} \epsilon_{z \nu \alpha \beta} A_{\nu} \partial_{\alpha} A_{\beta}. 
\eeq
In its primitive form above, Eq.~\eqref{eq:4} does not actually look like a topological term, since it explicitly contains a preferred direction in 
space ($z$) and depends on a nonuniversal microscopic lattice constant $d$ through the Weyl node separation $2 Q$. 

To fix these issues, it proves useful to introduce the concept of a translation gauge field.~\cite{Volovik19,Nissinen21,Song21,Barkeshli21,Gioia21}
Recall that Bravais lattice points $\bR$ of a perfect crystal may be described as intersections of families of crystal planes, perpendicular to primitive 
reciprocal lattice vectors $\bb^i$, where $i = 1, 2, 3$ (or $x, y, z$ for a cubic crystal). 
Mathematically, this is expressed by the equation
\beq
\label{eq:5}
\theta^i (\br, t) = \bb^i \cdot \br = 2 \pi n^i, 
\eeq
where $n^i$ are sets of integers, labeling the crystal planes in a family $i$ and the Bravais lattice vectors $\br = \bR$ are the solutions of this equations. 
Eq.~\eqref{eq:5} implies that the reciprocal lattice vectors in a perfect crystal may be expressed as gradients of the phases $b^i_j = \partial_j \theta^i$. 
This may be generalized to a distorted crystal, including time-dependent distortions, by introducing translation ``gauge fields"
\beq
\label{eq:6}
e^i_{\mu} = \frac{1}{2 \pi} \partial_{\mu} \theta^i. 
\eeq
The fields $e^i_{\mu}$ may in fact be viewed as true (strictly speaking, integer valued) gauge fields, if one explicitly takes account of the fact that the phases 
$\theta^i$ on crystal planes may be relabelled in arbitrary $2 \pi \times \mathbb{Z}$  increments.~\cite{Song21,Barkeshli21} This will not make a significant difference in our case and either viewpoint is acceptable. 

In a convenient differential form language, we may view $e^i$ as a one-form 
\beq
\label{eq:7}
e^i = e^i_{\mu} d x^{\mu}. 
\eeq
In a crystal without dislocations, 
\beq
\label{eq:8}
d e^i = \frac{1}{2} (\partial_{\mu} e^i_{\nu} - \partial_{\nu} e^i_{\mu}) dx^{\mu} \wedge dx^{\nu} = 0, 
\eeq
as clearly follows from the definition Eq.~\eqref{eq:6}. 
On the other hand, if a dislocation with a Burgers vector along $\bb^i$ is present, the integral of $e^i$ around a loop, enclosing the dislocation line 
is $\oint e^i = 1$. 

The benefit of introducing translation gauge fields becomes apparent if we now replace a reciprocal lattice vector along the $z$-direction in Eq.~\eqref{eq:4} 
with the corresponding translation gauge field
\beq
\label{eq:9}
\frac{2 \pi}{d} \delta^z_{\mu} \ra 2 \pi e^z_{\mu}. 
\eeq
Then Eq.~\eqref{eq:4} becomes
\beq
\label{eq:10}
\cL = i \frac{\lambda}{4 \pi} \epsilon_{\mu \nu \alpha \beta} e^z_{\mu} A_{\nu} \partial_{\alpha} A_{\beta}
 = i \frac{\lambda}{4 \pi} e^z \wedge A \wedge d A, 
\eeq
where $\lambda = 2 Q/ (2 \pi/d)$ is a dimensionless separation between the Weyl nodes in units of the reciprocal lattice vector. 
Now Eq.~\eqref{eq:10} looks like a proper topological term, which only contains gauge fields and a universal coefficient. 
The nonuniversal and variable lattice constant $d$ has been absorbed into the definition of the translation gauge field and we will 
henceforth set $d = 1$ for simplicity. 
Since we are using the imaginary-time formulation, upper and lower indices do not need to be distinguished and we will use 
lower indices for space-time components of the gauge fields throughout. 
Varying Eq.~\eqref{eq:10} with respect to $e^z_z$ produces response per atomic $xy$-plane, which is determined by a universal 
numerical coefficient $\lambda$. 
A noninteger value of the coefficient $\lambda$ requires gapless modes in the form of a pair of Weyl nodes to be present,~\cite{Gioia21,Wang21}
since a fractional value (in units of $e^2/h$) of the Hall conductance per atomic plane is impossible in a noninteracting gapped insulator. 

\subsection{3D analog of the Pfaffian-antisemion state}
\label{sec:2.2}
To derive the field theory of the gapped 3D FQH state of Ref.~\onlinecite{Wang20} we first move to a dual description 
of the noninteracting Weyl semimetal of Eq.~\eqref{eq:1}, in which the electric charge is separated from the fermions
and is represented in terms of a two-form gauge potential, which couples to the vortex loop excitations.~\cite{Senthil08,Barkeshli12,Thakurathi20}
This approach is similar to what is known as ``functional bosonization",~\cite{Burgess94,Burgess94-2,Fradkin13,Rosenow13}
apart from unimportant technical details. 
We start by representing the fermion operators in Eq.~\eqref{eq:1} (after Fourier transforming them to real space) as
\beq
\label{eq:11}
\psi_r = e^{i \theta_r} f_r, 
\eeq
where $r$ label the sites of a cubic lattice, $e^{i \theta_r}$ represents a spinless charged boson (chargon) and  $f_r$ is a neutral fermion (spinon).  
After straightforward and standard manipulations,~\cite{Thakurathi20,Palumbo20,Grushin20} one obtains the following exact representation of the Weyl semimetal 
Lagrangian $\cL$
\beq
\label{eq:12}
\cL = \cL_f + \cL_b
\eeq
where 
$\cL_f$ is the Lagrangian of the spinons $f_r$, which has a form, identical to the lattice Lagrangian of the original electrons $\psi_r$, 
except that $f_r$ are coupled to a compact statistical gauge field $a_{\mu}$ rather than the probe electromagnetic field $A_{\mu}$. 
The statistical field expresses $U(1)$ gauge invariance of the parton decomposition Eq.~\eqref{eq:11} and serves the purpose of gluing together 
the spinons and the chargons. 
The chargon Lagrangian has the form 
\beq
\label{eq:13}
\cL_b = \frac{i}{4 \pi} (A_{\mu} - a_{\mu}) \epsilon_{\mu \nu \alpha \beta} \Delta_{\nu} b_{\alpha \beta} + \frac{1}{8 \pi^2 \chi} (\epsilon_{\mu \nu \alpha \beta}
\Delta_{\nu} b_{\alpha \beta})^2. 
\eeq
Here $b_{\mu \nu} = -b_{\nu \mu}$ is a two-form $2\pi \times \mathbb{Z}$ valued lattice gauge field, which represents integer chargon currents $J_{\mu}$ as
\beq
\label{eq:14}
J_{\mu} = \frac{1}{4 \pi} \epsilon_{\mu \nu \alpha \beta} \Delta_{\nu} b_{\alpha \beta}, 
\eeq 
$\Delta_{\mu}$ is a lattice derivative and $\chi$ is a positive constant. Lattice site indices $r$ have been suppressed everywhere for brevity. 

To avoid dealing with discrete variables, we may implement the $2 \pi \mathbb{Z}$ constraint on $b_{\mu \nu}$ by adding a term 
$\frac{i}{2} \tilde J_{\mu \nu} b_{\mu \nu}$  to $\cL_b$ and summing over integer-valued variables $\tilde J_{\mu \nu}$, which have the meaning of 
vortex loop currents. 
Gauge invariance of Eq.~\eqref{eq:14} with respect to a transformation $b_{\mu \nu} \ra b_{\mu \nu} + \Delta_{\mu} g_{\nu} - \Delta_{\nu} g_{\mu}$ 
implies a conservation law
\beq
\label{eq:15}
\Delta_{\mu} \tilde J_{\mu \nu} = 0, 
\eeq
which may be solved as 
\beq
\label{eq:16}
\tilde J_{\mu \nu} = \frac{1}{2 \pi} \epsilon_{\mu \nu \alpha \beta} \Delta_{\alpha} c_{\beta},
\eeq
where $c_{\mu}$ are $2 \pi \mathbb{Z}$ valued one-form gauge fields. 
The constraint on $c_{\mu}$ may, in turn, be softened by adding a term $-t \cos(\Delta_{\mu} \phi + c_{\mu})$, where the presence of a new compact 
angular variable $\phi$ takes account of the gauge invariance of Eq.~\eqref{eq:16} with respect to $c_{\mu} \ra c_{\mu} + \Delta_{\mu} \phi$. 
In essence, the particle created by $e^{i \phi}$, is the original chargon. 

Then, after taking the continuum limit, the chargon Lagrangian takes the dual form 
\beq
\label{eq:17}
\cL_b = \frac{i}{4 \pi} (A_{\mu} - a_{\mu} + c_{\mu}) \epsilon_{\mu \nu \alpha \beta} \partial_{\nu} b_{\alpha \beta} + \ldots, 
\eeq
where $\ldots$ contain both the higher-derivative terms for $b_{\mu \nu}$ and the additional terms for $c_{\mu}$ whose form depends on the 
value of the parameter $t$. 
In particular, when $t$ is large, $e^{i \phi}$ boson is condensed, leading to a mass term for $c_{\mu}$ (i.e. gap for vortices), which may then be ignored. 
Integration over $b_{\mu \nu}$ the simply sets $A_{\mu} = a_{\mu}$, i.e. the electric charge is re-attached to the spinons and we recover the original 
Weyl semimetal. 
In contrast, when $t$ is small, $e^{i \phi}$ particle is gapped, which leads to a Maxwell term, $(\epsilon \partial c)^2$, for the gauge field $c_{\mu}$. 
In this case, integration over $c_{\mu}$ produces a mass term $b^2$ for the two-form gauge field, which corresponds to a charge gap. 
This state is a Mott insulator, which has gapless spinons that retain the Weyl semimetal band structure. 

To obtain a fully gapped state, which preserves topological response of the Weyl semimetal Eq.~\eqref{eq:10} and does not break any symmetries, 
we place the spinons into a paired state. 
For weak pairing, only the intra-nodal pairing state opens a gap.~\cite{Meng12,Moore12,Bednik15,YiLi18}
Such a pairing generally breaks translational symmetry, except when $2 Q = \pi$ or $\lambda = 1/2$,~\cite{Wang20} to which we now specialize. 
With such an intra-nodal pairing term, the spinon Hamiltonian may be brought to the form
\beqa
\label{eq:18}
&&H = \frac{1}{2} \sum_\bk f^\dg_\bk \left\{\sigma_x \sin(k_x) + \sigma_y \sin(k_y) \right. \nonumber \\
&+&\left.\left[\sqrt{\Delta^2 + \cos^2(k_z)} - \tilde m (2 - \cos(k_x) - \cos (k_y)) \right] \sigma_z \right\} f^\pdg_\bk, \nonumber \\
\eeqa
where $\Delta$ is the pairing amplitude. 
This Hamiltonian describes a 3D topological $p$-wave superconductor, which has a chiral Majorana mode, spanning the full extent of the BZ. 
This may also be viewed as a stack of 2D topological superconductors, since the pairing gap does not close at any value of $k_z$. 

The spinon pairing produces a term $\propto - \cos(2 a_{\mu})$ for the statistical gauge field, which leaves only $a_{\mu} = 0, \pi \,\textrm{mod}\, 2 \pi$ possible 
values at low energies and makes it a $\mathbb{Z}_2$ gauge field. 
While nontrivial $\pi$-flux configurations of $a_{\mu}$ (visons~\cite{SenthilFisher}) are still possible, these may be easily shown to bind gapless 1D Majorana mode in their 
cores, which is a direct consequence of the fact that the spinon superconductor is topologically nontrivial. 
This means that in any fully gapped symmetry-preserving state such vison loop excitations must be gapped, which means that we may set 
$a_{\mu} = 0 \,\textrm{mod } \,2 \pi$ at low energies. 
This detaches the boson and fermion sectors of the theory. The fermion sector thus contributes the same thermal Hall response as the noninteracting Weyl 
semimetal at $\lambda = 1/2$, which arises from the chiral Majorana mode, spanning the full BZ. 
The electrical response must entirely come from the boson sector. 

In order to reproduce the electrical response of the noninteracting Weyl semimetal, it is necessary to condense double (i.e. flux $4 \pi$) 
vortices of the boson field $e^{i \theta}$. This is accomplished by the following modification of the field theory Eq.~\eqref{eq:17}
\beqa
\label{eq:19}
\cL_b&=&\frac{i}{4 \pi} (A_{\mu} + 2 c_{\mu}) \epsilon_{\mu \nu \alpha \beta} \partial_{\nu} b_{\alpha \beta} + \frac{2 i}{4 \pi}
\epsilon_{\mu \nu \alpha \beta} e^z_{\mu} c_{\nu} \partial_{\alpha} c_{\beta} \nonumber \\
&+&\frac{1}{2 g} (\epsilon_{\mu \nu \alpha \beta} \partial_{\alpha} c_{\beta})^2 + \frac{i}{2} b_{\mu \nu} \tilde j_{\mu \nu} + i c_{\mu} j_{\mu}.
\eeqa
The extra factor of $2$ in front of $c_{\mu}$, compared to Eq.~\eqref{eq:17}, expresses the fact that double (flux $4 \pi$) vortices are being condensed. 
This also means that the quasiparticle, which is minimally coupled to the gauge field $c_{\mu}$, carries a charge $1/2$. 
The second term is a topological term, which will give rise to the correct electrical Hall conductivity, as will be shown below. 
This term may be viewed as describing a layered integer quantum Hall state of the charge-$1/2$ bosonic quasiparticles. 
The third term is the Maxwell term. It is subdominant to the topological term at long wavelengths, but has been included explicitly since the topological term 
only contains components of $c_{\mu}$, transverse to the translation gauge field $e^z$. In particular, if $e^z_{\mu} = \delta^z_{\mu}$, $c_z$ does not enter into the topological term and its dynamics is governed by the Maxwell term. Finally, $j_{\mu}$ and $\tilde j_{\mu \nu}$ represent bosonic quasiparticle and vortex source currents, which are minimally coupled to the gauge fields $c_{\mu}$ and $b_{\mu \nu}$ correspondingly. 

 Let us now demonstrate that Eq.~\eqref{eq:19} indeed reproduces topological response of a noninteracting Weyl semimetal.  Let us set $\tilde j_{\mu \nu} = 0$ and integrate out $b_{\mu \nu}$. 
 This gives $c_{\mu} = - A_{\mu}/2$.  Substituting this back into Eq.~\eqref{eq:19}, we obtain
 \beq
 \label{eq:20}
 \cL_b = \frac{i}{8 \pi} \epsilon_{\mu \nu \alpha \beta} e^z_{\mu} A_{\nu} \partial_{\alpha} A_{\beta} - \frac{i}{2} A_{\mu} j_{\mu}. 
 \eeq
 The first term in Eq.~\eqref{eq:20} correctly reproduces the electrical Hall conductivity of a noninteracting Weyl semimetal with $\lambda = 1/2$, 
 which is half conductance quantum $\sigma_{xy} = 1/4 \pi$ per atomic plane. 
 The second term tells us that quasiparticle excitations in the gapped state, described by Eq.~\eqref{eq:19}, are bosons with electric charge $1/2$. 
 To establish gapped nature of this state it is important to note the following. If we reinsert the statistical gauge field $a_{\mu}$ into Eq.~\eqref{eq:19}, 
 it is clear that fluctuations of $b_{\mu \nu}$ effectively constrain $c_{\mu} = (a_{\mu} - A_{\mu})/2$. This implies that, since $a_{\mu}$ is made a
 $\mathbb{Z}_2$ gauge field by spinon pairing, $c_{\mu}$ becomes a discrete $\mathbb{Z}_4$ gauge field. 
 This is important, since, unlike in $2+1$ dimensions, a $(3+1)$-dimensional Maxwell-Chern-Simons theory with $U(1)$ gauge fields is 
 gapless.~\cite{Carroll90,Ying23}
 
The most straightforward way to see that this theory also correctly captures the thermal Hall conductivity $\kappa_{xy} = 0$ is to consider the boundary
theory, that corresponds to Eq.~\eqref{eq:19}. 
To derive the boundary theory we follow the standard method.~\cite{Wen_QFT}
We choose a gauge, in which on the boundary, taken to be in the $xz$-plane,  we set the temporal components of all the gauge fields to zero, i.e. $c_0 = 0,\, b_{0 \mu} = 0$. 
Then, integrating out $c_0$, we obtain
\beq
\label{eq:21}
\epsilon_{0 \nu \lambda \rho} \partial_{\nu} b_{\lambda \rho} = \epsilon_{0 \nu \lambda \rho} \partial_{\nu} (e^z_{\lambda} c_{\rho} - e^z_{\rho} c_{\lambda}),
\eeq
while integrating $b_{0 \nu}$ gives
\beq
\label{eq:22}
\epsilon_{0 \nu \lambda \rho} \partial_{\lambda} c_{\rho} = 0. 
\eeq
Eqs.~\eqref{eq:21} and \eqref{eq:22} along with $d e^z = 0$ imply that 
\beq
\label{eq:23}
\epsilon_{0 \nu \lambda \rho} \partial_{\nu} b_{\lambda \rho} = 0. 
\eeq
Eq.~\eqref{eq:23} may then be solved as
\beq
\label{eq:24}
b_{i j} = \partial_i g_j - \partial_j g_i, 
\eeq
where $i, j = x, z$ refer to spatial coordinates on the boundary, while Eq.~\eqref{eq:22} is solved as 
\beq
\label{eq:25} 
c_i = \partial_i \varphi. 
\eeq
Plugging this back into what remains of Eq.~\eqref{eq:19} after integrating out $c_0$ and $b_{0 \mu}$, we obtain
\beq
\label{eq:26}
\cL_b = \frac{i}{2 \pi} \epsilon_{0 \nu \lambda \rho} e^z_{\nu} \partial_{\lambda} \varphi \partial_{\tau} \partial_{\rho} \varphi
- \frac{i}{\pi} \epsilon_{0 \nu \lambda \rho} \partial_{\nu} \varphi \partial_{\tau} \partial_{\lambda} g_{\rho}, 
\eeq
where $\partial_{\tau} \equiv \partial_0$. 
Integrating this in the presence of a boundary, perpendicular to the $y$-direction, gives
\beq
\label{eq:29}
\cL_{surf} = \frac{i}{2 \pi} \epsilon_{i j} e^z_i \partial_{\tau} \varphi \partial_j \varphi - \frac{i}{\pi} \epsilon_{i j} \partial_{\tau} \varphi \partial_i g_j, 
\eeq
where $i, j = x, z$. 
Adding symmetry-allowed non-topological terms and the electromagnetic field, we finally obtain the following surface state Lagrangian
\beqa
\label{eq:30}
\cL_{surf}&=&\frac{i}{2 \pi} \epsilon_{i j} e^z_i \partial_{\tau} \varphi \partial_j \varphi - \frac{i}{\pi} \epsilon_{i j} \partial_{\tau} \varphi \partial_i g_j 
+ \frac{v_{\varphi}}{2 \pi} (\partial_i \varphi)^2 \nonumber \\
&+&\frac{v_g}{2 \pi} (\partial_i g_j - \partial_j g_i)^2 + \frac{i}{2 \pi} \epsilon_{\mu \nu \lambda} A_{\mu} \partial_{\nu} g_{\lambda}. 
\eeqa 
Setting $e^z_{\mu} = \delta^z_{\mu}$ and Fourier transforming, we obtain the following expression for the excitation spectrum of the surface modes
\beq
\label{eq:31}
\epsilon(\bk) = - \frac{v_g k_x}{2} + \sqrt{\left(\frac{v_g^2}{4} + v_g v_{\varphi}\right) k_x^2 + v_g v_{\varphi} k_z^2}. 
\eeq
This looks like an ordinary anisotropic 2D superfluid dispersion, except for a ``tilt" in the $x$-direction due to the first term. 
However, the dispersion is still nonchiral, since there is always a pair of left- and right-handed modes for every value of $k_z$. 
Consequently, a straightforward calculation gives a vanishing thermal Hall conductivity in this state
\beq
\label{eq:32}
\kappa_{xy} \sim \int d k_x d k_z v_x(\bk) \epsilon(\bk) \frac{\partial n_B[\epsilon(\bk)]}{\partial T} = 0, 
\eeq
where $v_x(\bk) = \frac{\partial \epsilon(\bk)}{\partial k_x}$ and $n_B(\epsilon)$ is the Bose-Einstein distribution. 
The integral over $k_x$ in Eq.~\eqref{eq:32} vanishes since the left-handed ($k_x < 0$) and right-handed ($k_x > 0$) modes 
give a contribution that is equal in magnitude but opposite in sign. 

By construction, this state is a fully gapped symmetric state, which has an identical topological response to a noninteracting Weyl semimetal 
at $\lambda = 1/2$. 
Note again that, while there does exist a close connection between this state and the 2D Pfaffian-antisemion state, it may not be viewed as a simple 
stack of such 2D states. In particular, there are no semion quasiparticles, but isolated intersections of $2 \pi$ vortex loop excitations with atomic $xy$-planes 
do behave as semions. 

\subsection{3D analog of the PH-Pfaffian state}
\label{sec:2.3}
Now we note that a second distinct gapped symmetric state, reproducing topological response of a noninteracting Weyl semimetal, actually exists. 
This state is, in a way, simpler than the 3D analog of the Pfaffian-antisemion above and, as we will demonstrate, may be viewed as a 3D analog 
of the PH-Pfaffian.~\cite{Bonderson13,Fidkowski14,Son15,Son18}

To construct this state, we take a time-reversed copy of our Weyl semimetal with $\lambda = 1/2$. 
Writing its Lagrangian in terms of spinon and chargon variables, we have
\beq
\label{eq:33}
\cL = \bar f \gamma_{\mu} (\partial_{\mu} + i a_{\mu}) f - \frac{i}{8 \pi} e^z \wedge a \wedge d a + \frac{i}{4 \pi} ( A - a) \wedge d b, 
\eeq
where the first term is the contribution of the gapless Weyl fermions while the second term is the topological contribution from all the filled 
negative-energy states. We will switch to the index-free notation henceforth. 
We now place the chargons into a stack of independent $\nu = 1/2$ quantum Hall states in each $xy$-atomic plane. 
Technically, this means that we take the two-form gauge field $b$ to be ``foliated"~\cite{Slagle17,Slagle19,Slagle21,Karch21}
\beq
\label{eq:34}
b = e^z \wedge \tilde b, 
\eeq
where $\tilde b = \tilde b_0 d \tau + \tilde b_x d x + \tilde b_y dy$ is a one-form field that lacks the $z$-component, 
and add a term $- \frac{2 i}{4 \pi} e^z \wedge \tilde b \wedge d \tilde b$ to the Lagrangian Eq.~\eqref{eq:33}. 
Furthermore, we place the spinons into the intra-nodal pairing state of Eq.~\eqref{eq:18}, which leads to a 3D $p+ip$ topological superconductor with a chiral Majorana mode, spanning the surface BZ, whose chirality is, however, opposite to the chirality of the Fermi-arc state of the original noninteracting Weyl semimetal. 
This gaps out the gauge field $a_{\mu}$ and decouples the boson and fermion sectors. 

The boson sector Lagrangian now reads
\beq
\label{eq:35}
\cL_b = - \frac{2 i}{4 \pi} e^z \wedge \tilde b \wedge d \tilde  b + \frac{i}{2 \pi} e^z \wedge A \wedge d \tilde b. 
\eeq
Integrating over $\tilde b$ leaves the effective action
\beq
\label{eq:36}
\cL_b = \frac{i}{8 \pi} e^z \wedge A \wedge d A,
\eeq
which describes topological electrical response, which is identical to that of the original [i.e. not the time-reversed one of Eq.~\eqref{eq:33}] noninteracting 
Weyl semimetal, Eq.~\eqref{eq:10}. 
The thermal Hall effect, coming from $\cL_b$, is twice that of the noninteracting Weyl semimetal, however a minus a half is contributed by the 
opposite-chirality Majorana surface state of the paired time-reversed spinons. 
Thus we fully reproduce both electrical and thermal topological responses of the original noninteracting gapless Weyl semimetal. 

This state may be viewed as a 3D generalization of the 2D PH-Pfaffian state. 
Note that, unlike the 3D analog of the Pfaffian-antisemion state, described above, this state is not a 3D incompressible liquid,
but exhibits a fracton-type order.~\cite{Slagle17,Slagle19,Slagle21,Karch21}
If we ignore fermions, Eq.~\eqref{eq:35} describes a stack of independent 2D PH-Pfaffian states.
The charge-$1/2$ anyon excitations in these 2D states are only able to move within a given plane and can not tunnel between the planes. 
Neutral fermions propagate in 3D and connect the individual layers together, but the anyons remain confined within 2D layers. 
\section{``Dual" Weyl semimetal}
\label{sec:3}
The existence of a 3D analog of the PH-Pfaffian has an important implication, which we will now discuss. 
Let us first return back to the 3D Pfaffian-antisemion state. 
Let us note that, in this case, the topological response of a noninteracting Weyl semimetal is only reproduced when the fermionic spinons are 
gapped by pairing and vison vortex loops excitations are gapped. If the pairing gap is taken to zero, the statistical field $a$ is no longer massive 
and its coupling to the gapless Weyl spinons produces a topological term $\frac{i}{8 \pi} e^z \wedge a \wedge d a$, so that the Lagrangian
may be written as
\beqa
\label{eq:37}
\cL&=&\bar f \gamma_{\mu} (\partial_{\mu} + i a_{\mu}) f + \frac{i}{8 \pi} e^z \wedge a \wedge d a \nonumber \\
&+&\frac{i}{4 \pi} (A - a + 2 c) \wedge d b + \frac{i}{2 \pi} e^z \wedge c \wedge d c. 
\eeqa
Integrating out $b$ and $c$ gives 
\beqa
\label{eq:38}
\cL&=&\bar f \gamma_{\mu} (\partial_{\mu} + i a_{\mu}) f + \frac{i}{4 \pi} e^z \wedge a \wedge d a \nonumber \\
&-&\frac{i}{4 \pi} e^z \wedge A \wedge d a + \frac{i}{8 \pi} e^z \wedge A \wedge d A. 
\eeqa
To obtain the electromagnetic response, we now integrate out $a$. This may be done perturbatively, treating the response of the gapless low-energy modes, i.e. the first term in Eq.~\eqref{eq:38} as a perturbation, compared to the second term. 
This is possible, because the response of the gapless modes, treated in the random phase approximation (RPA), is given by
\beq
\label{eq:39}
S_f = \frac{1}{2} \sum_q a_{\mu}(q) \Pi_{\mu \nu}(q) a _{\nu} (-q),
\eeq
where 
\beq
\label{eq:40}
\Pi_{\mu \nu}(q) = (q^2 \delta_{\mu \nu} - q_{\mu} q_{\nu}) f(q^2), 
\eeq
is the polarization operator of the massless 3D Dirac fermion and 
\beq
\label{eq:41} 
f(q^2) = \frac{1}{12 \pi^2} \ln \left(\frac{4 \Lambda^2}{q^2} \right) + {\cal O}(1). 
\eeq
Here $\Lambda \gg q$ is the cutoff momentum, and a convention $q_0 = - \Omega$ is used ($\Omega$ is the Matsubara frequency). 
Note that $\Pi_{\mu \nu}(q)$ is almost the same as the polarization operator of the massive 3D Dirac fermion, in which case $f(q^2)$ would 
have been a constant at small $q$. Even with the log nonanalyticity, $\Pi_{\mu \nu}(q)$ is still much smaller, in the long wavelength limit, than the topological contributions, which are of first order in $q$. 

At leading order we may then ignore the gapless fermions and vary the Lagrangian with respect to $a$, which gives at the saddle point $a = A/2$ 
and leaves the Lagrangian
\beq
\label{eq:42}
\cL = \bar f \gamma_{\mu} (\partial_{\mu} + i A_{\mu}/2) f + \frac{i}{16 \pi} e^z \wedge A \wedge d A, 
\eeq
which clearly corresponds to half of the Hall conductivity of a noninteracting Weyl semimetal, i.e. the theory with gapless spinons does not reproduce 
topological response of the noninteracting Weyl semimetal. 

In contrast, let us return to Eq.~\eqref{eq:33}, which describes a time-reversed Weyl semimetal and add to it the foliated topological term of Eq.~\eqref{eq:35}, 
without opening the spinon pairing gap
\beqa
\label{eq:43}
\cL&=&\bar f \gamma_{\mu} (\partial_{\mu} + i a_{\mu}) f - \frac{i}{8 \pi} e^z \wedge a \wedge d a \nonumber \\
&+&\frac{i}{2 \pi} e^z \wedge (A - a) \wedge d \tilde b - \frac{2 i}{4 \pi} e^z \wedge \tilde b \wedge d \tilde  b. 
\eeqa
Integrating out $\tilde b$ now, we obtain
\beq
\label{eq:44}
\cL = \bar f \gamma_{\mu} (\partial_{\mu} + i a_{\mu}) f - \frac{i}{4 \pi} e^z \wedge A \wedge d a + \frac{i}{8 \pi} e^z \wedge A \wedge d A.
\eeq
This has identical electrical and thermal Hall responses to the original noninteracting Weyl semimetal. 
This means that Eq.~\eqref{eq:43} describes a distinct gapless state, which reproduces the topological response of a noninteracting 
Weyl semimetal. 
This statement is very closely analogous to the statement of duality between noninteracting 2D Dirac fermion and QED$_3$.~\cite{Wang-Senthil,Metlitski-Vishwanath,Son15,Alicea16,Seiberg16,Karch16,Raghu18} 
However, note that, in contrast to the 2D Dirac duality case, dynamically this system is quite different from a noninteracting Weyl semimetal. 
Indeed, integrating out $f$ and then $a$ in Eq.~\eqref{eq:44} using RPA produces a Meissner term for the components of $A$, transverse to $z$. 
The coefficient of the Meissner term, however, vanishes in the long-wavelength limit (it is equal to the inverse of the function $f(q^2)$, introduced in Eq.~\eqref{eq:41}). The system thus behaves as a superconductor at finite length scales and in directions, transverse to $z$, but with a phase stiffness that vanishes in the thermodynamic limit. In contrast, it behaves as an insulator along $z$. 
\section{Topological order in a gapped nodal line semimetal}
\label{sec:4}
We will now extend the ideas, developed above, to the case of nodal line semimetals, which realize a distinct kind of $(3+1)$-dimensional topological order, when gapped without breaking the protecting symmetries. 
In the nodal line semimetals, only nodal lines which arise from touchings of pairs of nondegenerate bands, are topologically nontrivial. 
In this case, TR symmetry may be taken to be broken, while the nodal line is then protected by the mirror reflection symmetry in the plane, 
containing the line.~\cite{Burkov11-2}
This may be described by the following two-band cubic-lattice Hamiltonian~\cite{Wang17,Shapourian18}
\beqa
\label{eq:45}
{\cal H}(\bk)&=&\left[6 - t_1 - 2 (\cos k_x + \cos k_y + \cos k_z) \right] \sigma_x \nonumber \\
&+&2 t_2 \sin k_z \sigma_y. 
\eeqa
The nodal line in this model appears in the $xy$-plane of the momentum space and is protected by the mirror reflection symmetry within this plane, where 
the mirror reflection operator is $\sigma_x$. 
The band-touching line in the $xy$-plane is given by the solution of the equation
\beq
\label{eq:46}
4 - t_1 - 2 (\cos k_x + \cos k_y ) = 0.
\eeq

\begin{figure}[t]
\includegraphics[width=\linewidth]{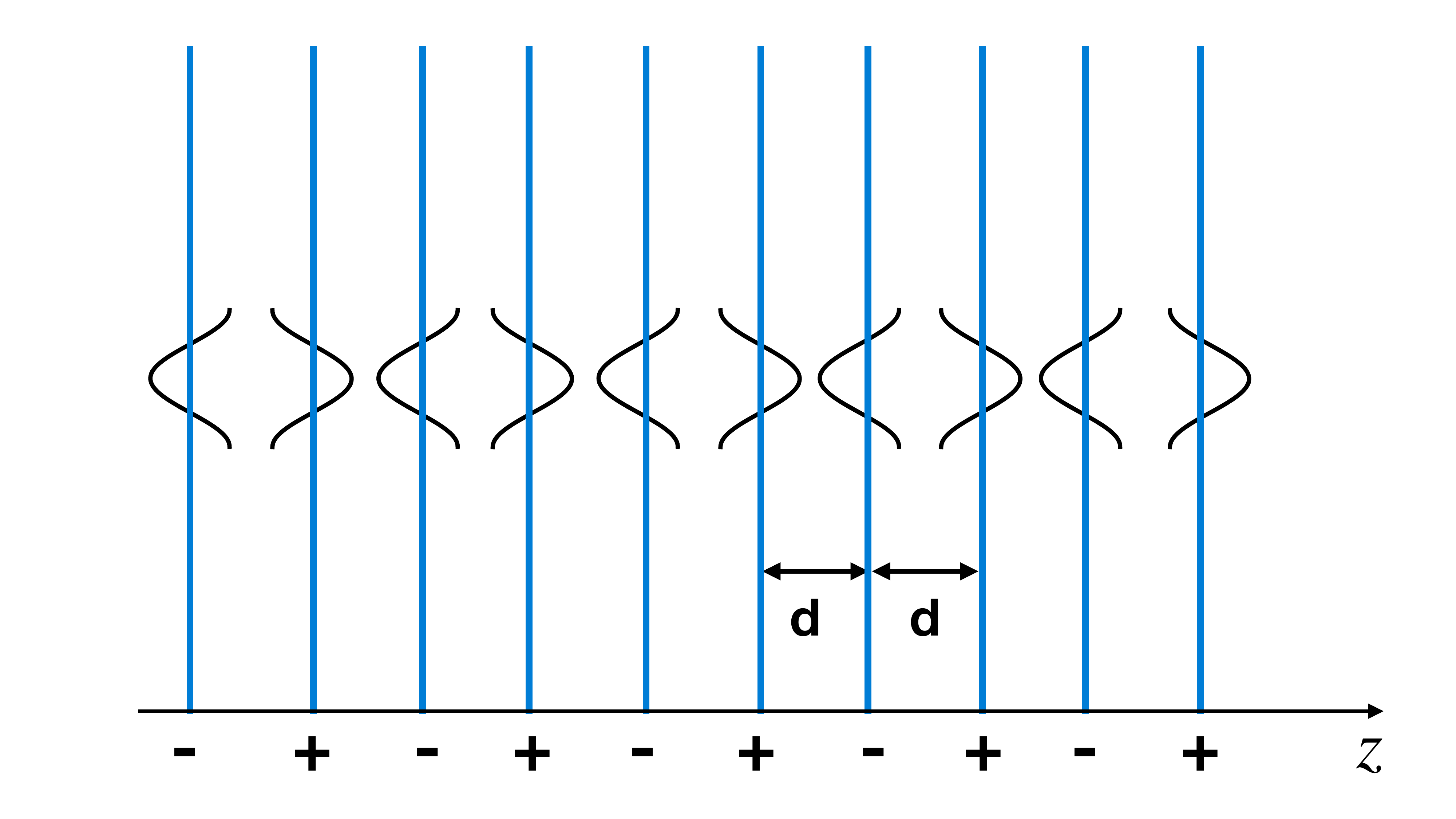}
\caption{(Color online) Construction of the nodal line semimetal as a stack ($z$ is the stacking direction) of alternating coupled electron and hole-like Fermi liquids,
indicated schematically by their dispersions. The Luttinger volume of each 2D Fermi liquid is equal in magnitude to the area in momentum space, enclosed by the nodal line. The lattice constant $d$ is set equal to unity in all the equations.}
\label{fig:1}
\end{figure}

In order construct a gapped symmetric state, it is useful to reinterpret Eq.~\eqref{eq:45} as a stacking of alternating electron and hole-like 
Fermi liquids with the band dispersions (see Fig.~\ref{fig:1})
\beq
\epsilon_{\pm}(\bk) = \pm [4 - t_1 - 2 (\cos k_x + \cos k_y )], 
\eeq
where $\pm$ are the two eigenvalues of the mirror reflection operator $\sigma_x$.~\cite{Wang21}
The Luttinger volumes of the two Fermi liquids $\pm V_F$ are equal in magnitude to the area in momentum space, enclosed by the nodal line. For the two Fermi liquids, the topological response describes the filling of the charged particles
\beq
\cL = \pm \frac{i V_F}{4 \pi^2} e^x \wedge e^y \wedge A, 
\eeq
Consequently, the topological response of the nodal line, takes the form of fractional electric polarization~\cite{Hughes_linenode,Burkov18-2,Wang21}
\beq
\label{eq:47}
\cL = \pm \frac{i V_F}{8 \pi^2} e^x \wedge e^y \wedge dA, 
\eeq
impossible in an ordinary insulator without topological order. 

The simplest way to obtain a gapped mirror-symmetric insulator with the same topological response Eq.~\eqref{eq:47} is to stack gapped 2D Fermi liquid states in a mirror-symmetric fashion. 
To gap the 2D Fermi liquids, we follow the same procedure as above. We represent an electron as a product of a neutral spinon $f$ and a bosonic 
chargon $e^{i \theta}$ and place the spinons into a fully-gapped paired state. 
The simplest fully gapped paired spinon state is $p$-wave (since the Fermi liquids are spinless), described by the following 
Hamiltonian
\beq
\label{eq:48}
H_f = \sum_{\bk} \left[\epsilon_{\pm}(\bk) f^\dg_{\bk}f^\pdg_{\bk} + \frac{\Delta}{2} (\sin k_x + i \sin k_y) f^\dg_{\bk} f^\dg_{-\bk} + \text{h.c.}\right]. 
\eeq
Introducing Nambu spinor notation $\psi_{\bk} = (f^\pdg_\bk, f^\dg_{- \bk})$, this may be represented as a massive 2D Dirac Hamiltonian
\beq
\label{eq:49}
H = \frac{1}{2} \sum_\bk \psi^\dg_\bk \left[\epsilon_{\pm}(\bk) \tau_z + \Delta (\tau_x \sin k_x - \tau_y \sin k_y) \right] \psi^\pdg_\bk, 
\eeq
where $\tau_a$ are Pauli matrices in the particle-hole space. 
This is the Hamiltonian of a Read-Green topological superconductor,~\cite{Read00}
which hosts chiral Majorana modes at the edges, with opposite chirality for electron and hole-like Fermi liquid states. 
Consequently, an elementary flux $h c/2 e = \pi$ vortex hosts a zero-energy localized Majorana bound state and can not be condensed.

To condense higher-flux vortices, we need to consider the chargon sector of the theory. 
Suppose we attempt to condense flux-$2 \pi$ vortices. The chargon sector will be described by the following theory~\cite{Balents05,Burkov05}
\beq
\label{eq:50}
\cL_b = \frac{i}{2 \pi} (A + c) \wedge d b \pm \frac{i V_F}{(2 \pi)^2} e^x \wedge e^y \wedge c. 
\eeq
Here $b$ is a one-form gauge field, which determines the charge current
\beq
\label{eq:51}
J_{\mu} = \frac{1}{2 \pi} \epsilon_{\mu \nu \lambda} \partial_{\nu} b_{\lambda}, 
\eeq
while $c$ is a one-form gauge field, which determines the vortex current
\beq
\label{eq:52}
\tilde J_{\mu} = \frac{1}{2 \pi} \epsilon_{\mu \nu \lambda} \partial_{\nu} c_{\lambda}.
\eeq
The last term of the Lagrangian produces the correct electromagnetic response of a system with charge $\nu = \pm V_F /(2 \pi)^2$ per unit cell when $b$ is integrated 
out, setting $c = -A$. 
However, when the filling $\nu$ is not an integer, Eq.~\eqref{eq:50} can not be the correct theory of a featureless 
insulator since the last term is not gauge invariant.
With $\nu = \pm p/q$, a featureless insulator may be obtained only by condensing flux $2 \pi q$ vortices, which is described by the theory
\beq
\label{eq:53}
\cL_b = \frac{i}{2 \pi} (A + q c) \wedge d b \pm i p e^x \wedge e^y \wedge c, 
\eeq
where all terms now have properly quantized integer coefficients and are gauge invariant. 
This is because the quasiparticle, minimally coupled to $c_{\mu}$, carries charge $1/q$, as seen from the first term. 
Therefore, the filling of such quasiparticles is $q \nu = p$ (i.e. an integer), which is what the second term expresses. 

Stacking such insulators with alternating sign of $\nu$ in the $z$-direction in a mirror-symmetric fashion, we obtain
\beq
\label{eq:54}
\cL_b = \frac{i}{4 \pi} (A + q c) \wedge d b \pm \frac{i p}{2} e^x \wedge e^y \wedge d c, 
\eeq
where the factor of $1/2$ in front of the last term arises due to the fact that the unit cell of the stack contains a pair of electron and hole-like 2D 
Fermi liquids and the mirror symmetry requires that all neighboring 2D Fermi liquids in the stack are separated by an equal distance. 
The gauge field $b$ in Eq.~\eqref{eq:54} has now been promoted to a two-form field, such that the $(3+1)$-dimensional charge current is given by
\beq
\label{eq:55}
J_{\mu} = \frac{1}{4 \pi} \epsilon_{\mu \nu \alpha \beta} \partial_{\nu} b_{\alpha \beta}, 
\eeq
while the two-form vortex current is 
\beq
\label{eq:56}
\tilde J_{\mu \nu} = \frac{1}{2 \pi} \epsilon_{\mu \nu \alpha \beta} \partial_{\alpha} c_{\beta}. 
\eeq
Integrating out $b$ in Eq.~\eqref{eq:54} gives $c = - A/q$ and the electromagnetic response described by 
\beq
\label{eq:57}
\cL = \pm \frac{i p}{2 q} e^x \wedge e^y \wedge d A = \pm \frac{i V_F}{8 \pi^2} e^x \wedge e^y \wedge d A, 
\eeq
which coincides with Eq.~\eqref{eq:47}.
Thus we obtain a featureless insulator with topological order, which has an identical topological response to a weakly-interacting nodal line semimetal.
Note that the nodal line semimetal has no topological thermal response, which is also the case in the fractionalized insulator that we have constructed. 
\section{Discussion and conclusions}
\label{sec:5}
In this paper we have presented a theory of $(3+1)$-dimensional topologically ordered states, obtained by gapping 3D topological semimetals without breaking protecting symmetries. 
We started by pointing out that a second gapped symmetric topologically-ordered state, preserving the chiral anomaly of magnetic Weyl semimetals,
exists, in addition to the state, originally proposed in Ref.~\onlinecite{Wang20}. 
We have shown that, while the state of Ref.~\onlinecite{Wang20} may be viewed as a 3D TR-breaking analog of the Pfaffian-antisemion
state in gapped 3D TI surface states, the new state is the 3D analog of the PH-Pfaffian. 
In contrast to the 3D Pfaffian-antisemion state, the 3D PH-Pfaffian does not exhibit a true 3D topological order, but a fracton-like order instead,
with independent layers of 2D PH-Pfaffian liquid immersed in a 3D $p + ip$ topological superconductor of neutral composite fermions. 

We then demonstrated that an interesting consequence of the existence of the 3D PH-Pfaffian, is a duality relation between a noninteracting
Weyl semimetal and QED$_4$, in which a time-reversed electrically-neutral Weyl semimetal is coupled to a dynamical gauge field, whose topological 
defects (intersections of flux lines with atomic planes) carry the electric charge. This duality relation may be viewed as a 3D generalization of the known Dirac fermion to QED$_3$ duality relation, but is weaker than in the 2D case, since the duality only applies to the topological response and not to the dynamics. 

Finally, we have extended the theory to include topological orders in a gapped nodal line semimetal. Other extensions, in particular to TR-invariant Weyl and 
Dirac semimetals are also possible, but do not lead to any fundamentally new structure. 
One lesson we may highlight is that gapped symmetric topological semimetals provide a very simple and natural setting for $(3+1)$-dimensional 
topologically-ordered states to appear. The simplicity stems, in part, from the fact that, due to the existence of a preferred direction, selected by either the separation between the Weyl points in momentum space, or the plane of the nodal line, there exists a natural connection to well-studied $(2+1)$-dimensional topological orders. 
The connection manifests either directly, in the form of layered fracton-like order, or less directly, when intersections of $(3+1)$-dimensional vortex-loop excitations with atomic planes behave as fractionally-charged and sometimes anyonic $(2+1)$-dimensional quasiparticle excitations. 

\begin{acknowledgments}
We acknowledge useful discussions with Chong Wang. Financial support was provided by the Natural Sciences and Engineering Research Council (NSERC) of Canada.
AAB was also supported by Center for Advancement of Topological Semimetals, an Energy Frontier Research Center funded by the U.S. Department of Energy Office of Science, Office of Basic Energy Sciences, through the Ames Laboratory under
contract DE-AC02-07CH11358. 
Research at Perimeter Institute is supported in part by the Government of Canada through the Department of Innovation, Science and Economic Development and by the Province of Ontario through the Ministry of Economic Development, Job Creation and Trade.
\end{acknowledgments}
\bibliography{references}
\end{document}